\documentclass[showpacs,twocolumn,aps,superscriptaddress]{revtex4}
\usepackage[applemac]{inputenc}
\usepackage{amssymb,amsmath}
\usepackage{color}
\usepackage{graphicx} 
\usepackage{epstopdf} 
\usepackage{enumerate}
\begin{document}
\title{Mass gap in the critical gravitational collapse of a kink}
\author{W. Barreto}
\affiliation{Centro de F\'\i sica Fundamental, Universidad de Los Andes, M\'erida 5101, Venezuela}
\affiliation{Departamento de F\'\i sica Te\'orica, Instituto de F\'\i sica A. D. Tavares,
Universidade do Estado do Rio de Janeiro, R. S\~ao Francisco Xavier, 524,
Rio de Janeiro 20550-013, RJ, Brasil}
\author{J. A. Crespo}
\affiliation{Departamento de F\'\i sica Te\'orica, Instituto de F\'\i sica A. D. Tavares,
Universidade do Estado do Rio de Janeiro, R. S\~ao Francisco Xavier, 524,
Rio de Janeiro 20550-013, RJ, Brasil}
\author{H. P. de Oliveira}
\affiliation{Departamento de F\'\i sica Te\'orica, Instituto de F\'\i sica A. D. Tavares,
Universidade do Estado do Rio de Janeiro, R. S\~ao Francisco Xavier, 524,
Rio de Janeiro 20550-013, RJ, Brasil}
\author{E. L. Rodrigues}
\affiliation{Departamento de Ci\^encias Naturais, Instituto de Bioci\^encias,
Universidade Federal do Estado do Rio de Janeiro,\\
Av. Pasteur, 458, Rio de Janeiro 22290-040, RJ, Brasil} 
\author{B. Rodriguez-Mueller}
\affiliation{Computational Science Research Center, San Diego State University, United States of America} 
\date{\today}
\begin{abstract}
We study the gravitational collapse of a kink within spherical symmetry and the characteristic formulation of General Relativity. We explore some expected but elusive gravitational collapse issues which have not been studied before in detail, finding new features. 
{The numerical one-parametric solution and the structure of the spacetime  {are calculated} using finite differences, Galerkin collocation techniques,
and some scripting for automated grid coverage.} {We study the threshold of black hole formation and {confirm} a mass gap in the phase transition.} In the supercritical case we find a mass scaling power law  $M_{BH}={M^*_{BH}}+K[\lambda-\lambda^*]^{2\gamma}+f(K[\lambda-\lambda^*]^{2\gamma})$, with $\gamma\approx 0.37$ independent of the initial data for the cases considered, and $M^*_{BH}$, $K$ and $\lambda^*$ each depending on the initial datum. The spacetime has a self-similar structure with a period of $\Delta\approx 3.4$. In the subcritical case the Bondi mass at null infinity decays in cascade with $\Delta/2$ interval as expected.
\end{abstract}
\maketitle
\section{Introduction} 
The practical limit of General Relativity as it nears the threshold of the quantum realm is an open issue.
{The answer might involve elucidating where relativity fails and whether there is a place where both theories merge.}
{In the strong field limit near the formation of a black hole Choptuik, discovered critical behavior \cite{c93} for a massless}
scalar field under spherical symmetry and minimally coupled to gravity. Choptuik found: (i) a critical behavior of Type II with a very small black hole mass; 
{(ii) an unstable naked singularity by fine-tuning generic initial data};   
{(iii) that the system follows a power law mass scaling and shows discrete self-similarity.}
Critical behavior of Type I is found when a massive scalar field (a Compton wavelength) is considered \cite{brady}, \cite{ss}.
For a review on the critical phenomena for gravitational collapse, including quantum extensions, see Ref. \cite{gm}.

{A self-gravitating massless scalar field has been a pivotal toy model for Numerical General Relativity in many different ways: Astrophysically, in the black hole coalescence and merging to predict expected profiles of gravitational radiation, relevant for detection \cite{pretorius}; Cosmologically, a possible connection with dark matter has been speculated (see Ref. \cite{darkm} and references therein); Mathematically, the field equations of the Einstein-Klein- Gordon system under spherical symmetry have the same structure of those off spherical symmetry in vaccum \cite{winicour}.}

Although the critical behavior was studied and comprehended deeply, the massless scalar field still remains a good playground to explore the non linear nature of gravity. In the critical behavior of Type II, is the final mass finite? \cite{psa05}. 
When the spacetime contains a black hole, power law scaling with a mass gap is not obvious \cite{mc}, \cite{cd15}. What about if we collapse a kink \cite{bglw96} up to a black hole formation? From physical grounds the critical behavior should not depend in general on boundary-initial conditions.
{We describe this system in terms of radiation coordinates \cite{bondi62,sachs62}, which in the case of spherical symmetry the line element takes the form \cite{bondi64}
\begin{equation}
    ds^2= e^{2\beta}du\left({Vr^{-1}}du+2dr\right)
    - r^2(d\theta^2 +\sin^2\theta d\phi^2). \label{eq:metric}
\end{equation}
where $\beta$ and $V$ are functions of $u$ and $r$. 
Here $u$ is a timelike coordinate; in a flat spacetime $u$ is just the retarded time. Therefore, surfaces $u=$ constant represent null cones open to the future; $r$ is a null coordinate ($g_{rr}=0$) such that surfaces $r=$ constant are spheres; $\theta$ and $\phi$ are the usual angular coordinates.}

In these coordinates, the Einstein-Klein-Gordon equations reduce
to~\cite{EKG,X1986}
\begin{equation}
     \beta_{,r}= 2\pi r(\Phi_{,r})^2    \label{eq:beta}
\end{equation}
\begin{equation}
     V_{,r}=e^{2\beta}             \label{eq:V}
\end{equation}
and the scalar wave equation $\Box \Phi = 0$, which takes the form
\begin{equation}
     2(r \Phi)_{,u r} = r^{-1}(rV\Phi_{,r})_{,r} . \label{eq:SWE}
\end{equation}
The initial null data necessary for evolution consists of
$\Phi(u_0,r)$, $r\ge R$, at initial time $u_0$. (We take $u_0 =0$). At
the mirror, we set $\Phi(u,R)=A=constant$, with the gauge condition
that $\Phi(u,\infty)=0$. We adopt the coordinate condition
$\beta(u,R)=0$.  The condition that the metric matches continuously to a
flat interior for $r<R$ requires $V(u,R)=R$.

With these conditions the scalar field and metric components have a
unique future evolution. The resulting metric does not have an asymptotic
Minkowski form at ${\cal I}^+$.  This is characterized by the quantity
$H(u)=\beta (u,\infty)$ which relates  Bondi time $t$ at ${\cal I}^+$
to the proper time $u$ at the reflecting boundary according to
$dt/d\tau = e^{2H}$. Bondi time is the physically relevant time for
distant observers.  The Bondi mass of the system can be expressed in
either an asymptotic or integral form~\cite{waveforms}:
\begin{eqnarray}
  M(u)&=&{1\over 2}e^{-2H}r^2({V\over r})_{,r}\,\Bigg| _{r=\infty}\nonumber \\
      &=& 2\pi\int_R^\infty e^{2(\beta-H)} r^2 (\Phi_{,r})^2 dr .
    \label{eq:mint}
\end{eqnarray}
\begin{figure}[!ht]
\begin{center}
\includegraphics[width=2.in,height=2.in]{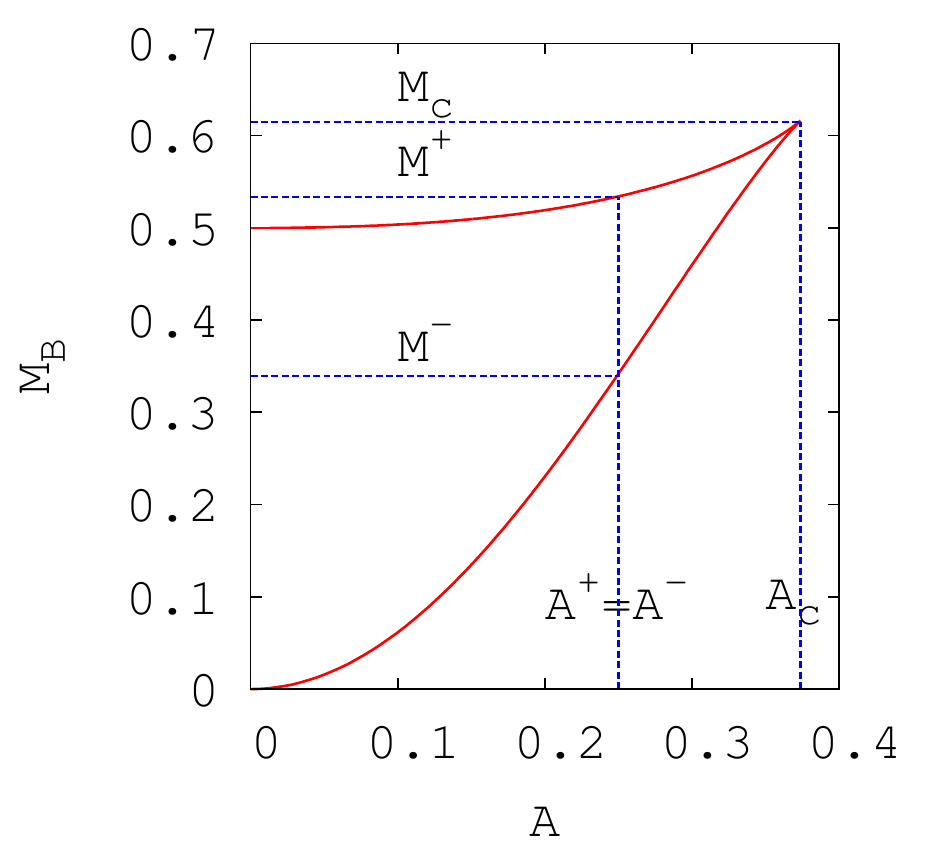}
\caption{The Bondi mass $M_{B}$ as a function of the amplitude $A$ for the kink formed with the JNW solution, with $R=1$. The critical solution $(A_c,M_c)$ without a mass gap is critical and refers to a turning point. Below $A_c$ the system has a mass gap $\Delta M=M^+-M^-$ which corresponds to the same amplitude $A^+=A^-$.
If we perturb the static solution $(A^-,M^-)$, below some critical parameter $\lambda^*$ the system decays to the static solution; beyond $\lambda^*$ the system always forms a black hole. If we perturb the static solution $(A^+,M^+)$, below some critical parameter $\lambda^*$ the system decays to the static solution with less energy; beyond $\lambda^*$ the system always forms a black hole. By construction of the kink we extract the naked singularity contained in the JNW solution when it is analytically extended to $r=0$.}
\label{twomass}
\end{center}
\end{figure}
\begin{figure}[!ht]
\begin{center}
\includegraphics[width=3.in,height=2.in]{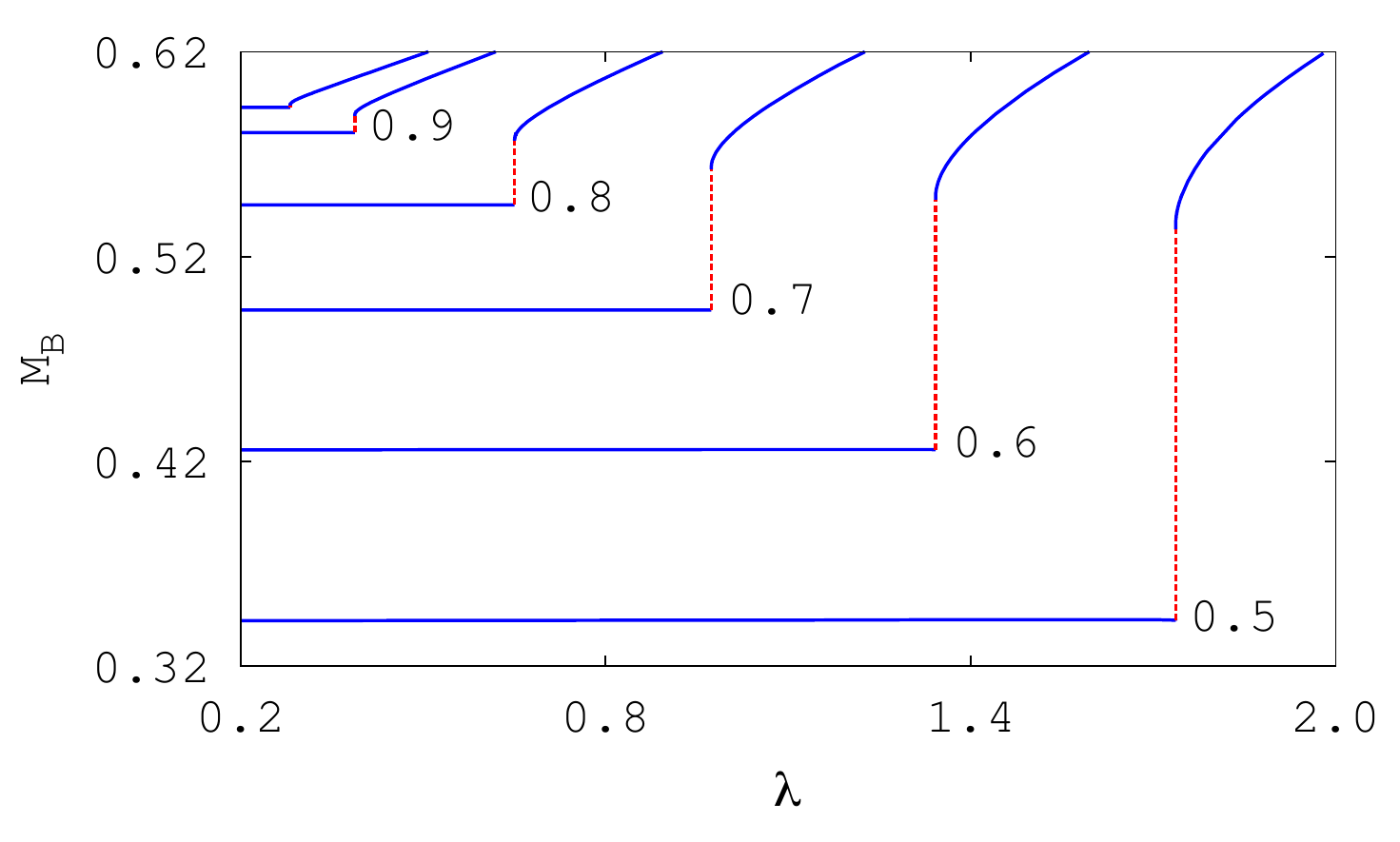}
\caption{The Bondi mass $M_{B}$ as a function of the parameter
$\lambda$ for different values of $\alpha$ in datum A: $0.5$; $0.6$; $0.7$; $0.8$; $0.9$ and $0.95$ (curve not labeled in graph). Near the critical value $\lambda^*$ for each $\alpha$ the mass gap is apparent, with a tendency to close near $\alpha\approx 0.95$. The grid size for these calculations is $N_x=3,072$.}
\label{massgapII}
\end{center}
\end{figure}
\begin{figure}[!ht]
\begin{center}
\includegraphics[width=3.in,height=2.in]{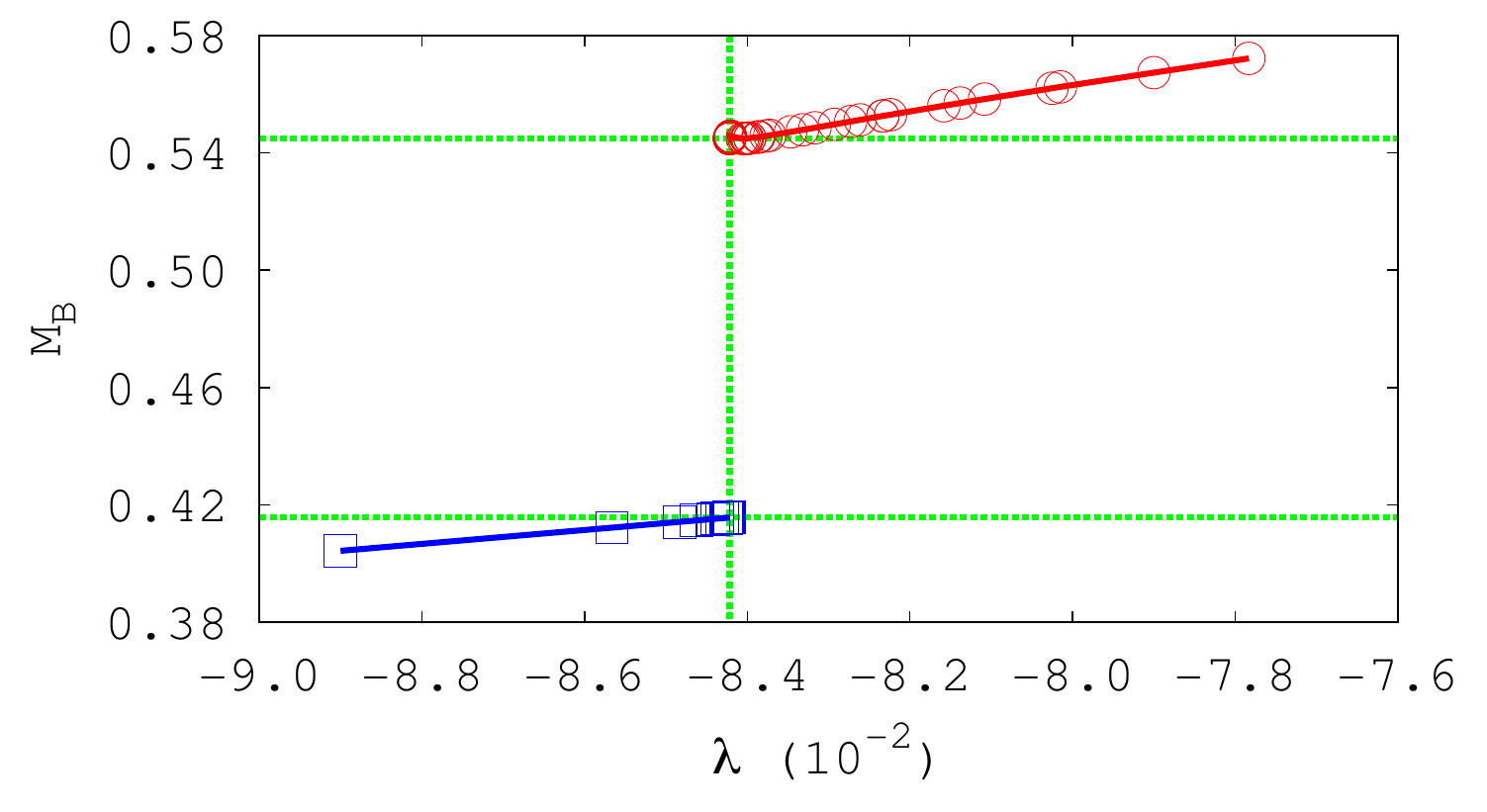}
\caption{The Bondi mass $M_{B}$ as a function of the parameter
$\lambda$. Near the critical value $\lambda^*$ the mass gap is apparent. This calculation corresponds to the initial datum D 
for $N_x=10^4$. Squared line represents the final Bondi mass for subcritical evolutions $\lambda < \lambda^*$ reaching the limit value of $M_{\lambda < \lambda^*}\approx 0.4157$; circled line represents the black hole mass for supercritical evolutions $\lambda^* < \lambda$ reaching the limit value of $M_{\lambda^* < \lambda}\approx 0.5450.$}
\label{massgap}
\end{center}
\end{figure}
\begin{figure}[!ht]
\begin{center}
\includegraphics[width=3in,height=2.in]{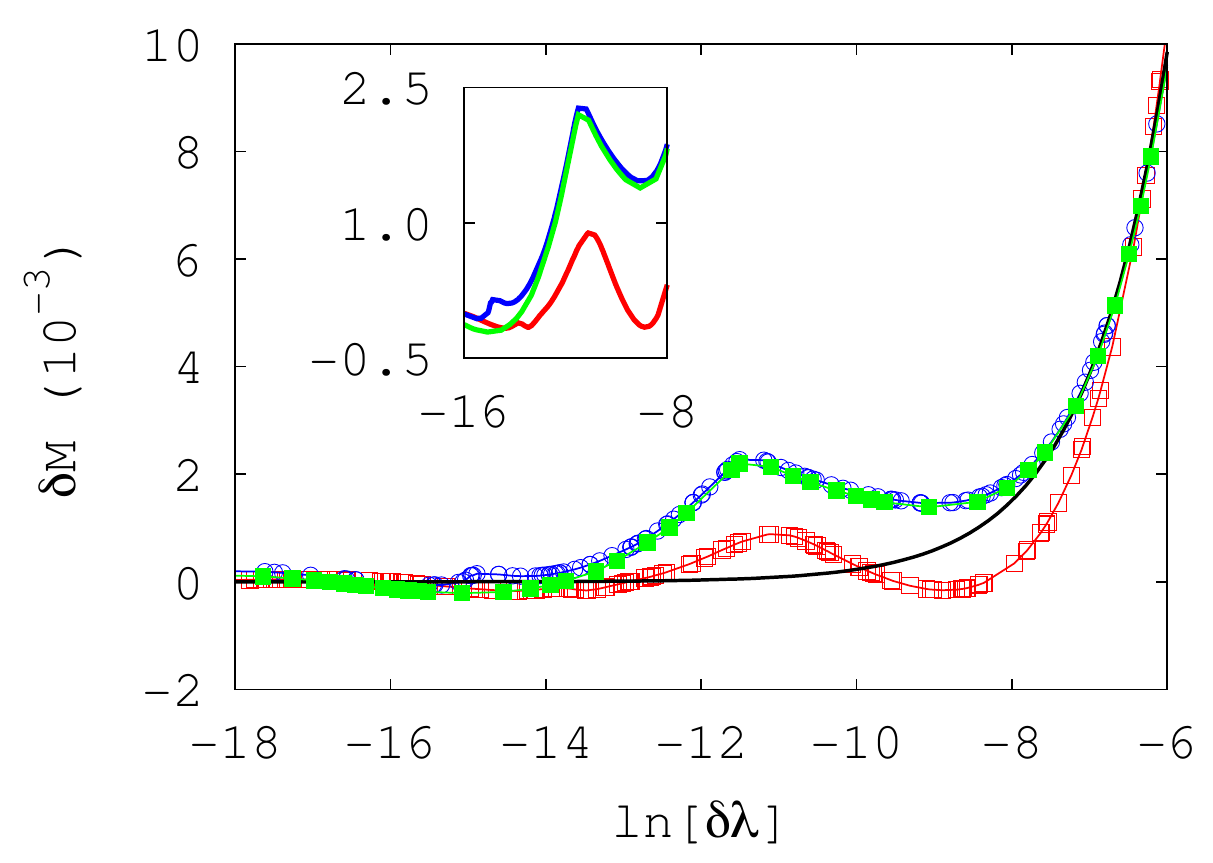}
\caption{$\delta M$ as a function of $\ln(\delta\lambda)$ for the initial data C (blue circles) and D (red squares), with $N_x=1.5\times 10^4$. The continuous line (black) roughly grows up with a power of $2\gamma\approx0.74$. From the oscillatory main component we get a period of $\Delta/2\gamma\approx 4.59$. The inserted graph shows the same functions in a narrow window. Other unstable modes of decaying oscillations are apparent. In this graph we also show results using the Galerkin-Collocation method for datum D (green squares) with $N=350$. Both methods, finite differences and Galerkin-Collocation give the same results up to some resolution.}
\label{critical}
\end{center}
\end{figure}
\begin{figure}[!ht]
\begin{center}
\includegraphics[width=3in,height=2.in]{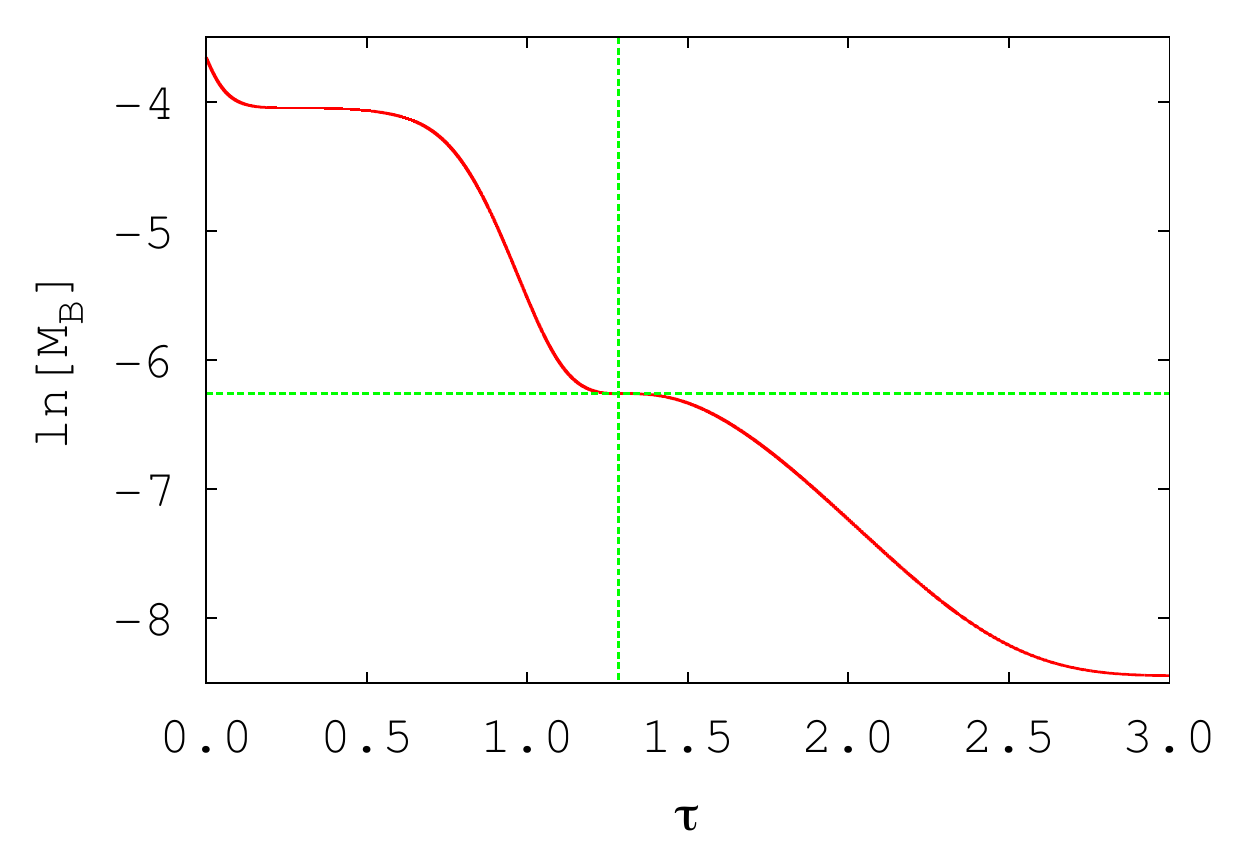}
\caption{$\ln[M_B]$ as a function of $\tau$ for the initial datum D, with $N_x=1.5\times 10^4$. The mass decays in the lower right quadrant roughly with a period of $\Delta/2$ as expected.}
\label{mass}
\end{center}
\end{figure}
\section{Static and other kinks}
The asymptotically flat static solution $\Psi$ of the
Einstein-Klein-Gordon system is an extremum of the energy, subject to a
fixed kink potential \cite{bglw96}.  This solution~\cite{jnw1968}, of Janis-Newman-Winicour (JNW), can be obtained
in null coordinates by setting $\Phi_{,u}=0$ in the wave
equation~(\ref{eq:SWE}). This gives
\begin{equation}
	  rV \Psi_{,r} = \text{const.},  \label{eq:lstat}
\end{equation}
whose solution, after using (\ref{eq:beta}) and (\ref{eq:V}) to
eliminate the $r$-dependence, is
\begin{equation}
\Psi(V) = {1 \over {4\sqrt{\pi} \>{\cosh} \alpha}} \ln
        \left[{{V + R\> (e^{2\>\alpha} - 1)} \over
         {V + R\> (e^{-2\>\alpha} - 1)}} \right] ,\label{eq:static}
\end{equation}
with $r(V)=r_\Psi$ given by
\begin{eqnarray}
        r_\Psi^2 &=&  e^{-4\> \alpha \tanh \alpha}
               [V + R\> (e^{-2\>\alpha} - 1)]^{1\> -\> \tanh \alpha}\times
               \nonumber\\
               &&\;\;\;\;\;\;\;\;\;\;\;\;\;\;\;\;\;[V + R\> (e^{2\>\alpha} - 1) ]^{1\> +\> \tanh \alpha}.
                \label{eq:rstatic}
\end{eqnarray}
Here the integration constant $\alpha$ determines
the kink potential,
\begin{equation}
       A_\Psi(\alpha)={\alpha \over \sqrt{\pi}\cosh\alpha}. \label{eq:kink}
\end{equation}
The spacetime has a naked singularity when analytically extended to
$r=0$~\cite{jnw1968}. The Bondi mass of this solution is
\begin{equation}
    M_{\Psi}(\alpha)=2R\sinh^2\alpha e^{-2\alpha\tanh\alpha}.
             \label{eq:statm}
\end{equation}
The one-parameter family of static equilibria has kink potential
$A_\Psi(\alpha)$, given by  (\ref{eq:kink}), which increases
monotonically with $\alpha$ from $A_\Psi(0)=0$ until it reaches a
maximum at the turning point $\alpha_c\approx 1.199$ satisfying
$\alpha_c\tanh\alpha_c=1$.  Above $\alpha_c$, $A_\Psi(\alpha)$
monotonically decreases to $0$ as $\alpha\rightarrow\infty$. Thus,
below $A_\Psi(\alpha_c)=A_c$, there are two static equilibria for each kink
amplitude. Similarly, the mass $M_\Psi(\alpha)$ increases
monotonically from $M_\Psi(0)=0$ to a maximum at the same turning point
$\alpha_c$ and then decreases monotonically to the black hole limit,
$M_\Psi(\alpha)\rightarrow R/2$, as $\alpha\rightarrow\infty$ \cite{bglw96}.
Figure 1 shows the basic setting of the system to study the critical behavior in the gravitational collapse of a kink. The static solution of JNW, and the physics behind it, is our starting point to build four types of kinks and study their dynamics in depth.
\subsection{Globally perturbed static solution}
From the JNW static solution, we construct a kink for any specific value of $\alpha$, truncating such a solution at $r=R$ and introducing the following simple global scale perturbation
\begin{equation}
\phi(0,r;\alpha)=\Psi + \lambda\left[ \frac{R-r}{2r(r+R)} \right]
\end{equation}
keeping the values of the static kink as boundary conditions. We know that for $\alpha=1$ and below $\lambda^*\approx0.1929$ the perturbed kink does not form a black hole; instead decays to the static solution. 
\subsection{Non compact initial kink}
For a kink potential $A>A_c$ no static equilibria exist. With these boundary conditions, we would expect any initial state to undergo collapse to a black hole \cite{bglw96}. We explore then this by considering initial data of the form
\begin{equation}
\phi(0,r)=\frac{2(A_c+\lambda)R}{(R+r)},
\end{equation}
for which a critical value of $\lambda^*=0$ no mass gap exist, and we have to wait an infinite proper time to observe whether or not a black hole is formed. 
\subsection{Compact initial kink}
From the above initial conditions and numerical experimentation we arrived to the following kink
\begin{equation}
\phi(0,r)=\frac{R(A_c+\lambda)}{re^{(r-R)^2/\sigma^2}},
\end{equation}
with a variance $\sigma=1/2$. The Gaussian-like shape makes this datum properly to explore the critical behavior in the gravitational collapse of the kink.
\subsection{Mixed initial kink}
In order to combine the features of data B and C we built the following initial datum, which resembles a Heaviside step-like descent function 
\begin{equation}
\phi(0,r)=\frac{R(A_c+\lambda)}{r}\frac{\{1+e^{(R-R_0)/\sigma}\}}{\{1+e^{(r-R_0)/\sigma}\}}
\end{equation}
with $R_0=10$ (step width related) and $\sigma=1/2$ (speed of descent related). These parameters let us adjust the dynamics and prescribe the formation or not of a black hole.
\section{Numerical methods, tests and other tools}
The problem has been studied using two different numerical methods 
and some scripting. We briefly resume here these well established numerical solvers and the wrapper scripts. 
\subsection{Finite differences}
We use a null cone evolution algorithm for nonlinear scalar waves developed in 
Refs. \cite{gwi,gw} (the 1D Pitt code) adapted to the present setting as reported in \cite{bglw96}.
The algorithm is based upon the compactified radial coordinate $x=r/(R+r)$, so that ${\cal I}^+$ is represented by a finite grid boundary, with $x=1/2$ at the mirror and $x=1$ at ${\cal I}^+$. The code has been tested to be globally second order accurate, i.e., the error in global quantities such as the Bondi mass is $O(x^2)$ in terms of the grid spacing $x$. This code has been used to get global energy conservation near the critical behavior \cite{b14} in a setting as originally studied by Choptuik. 

\subsection{Galerkin collocation}

We have performed numerical experiments using the Galerkin-Collocation method as described in Ref. \cite{cd15}. Briefly, we have introduced a new radial coordinate $\eta=r/R-1$ to place the reflector at $\eta=0$, and further compactified the domain $0 \leq \eta < \infty$ into $-1 \leq x < 1$. After introducing the auxiliary field $\Phi \equiv (1+\eta)\phi$, the relevant fields were approximated as series with respect to suitable basis functions that satisfy the boundary conditions. For instance,

\begin{equation}
\Phi(u,\eta) = \sum_{k=0}^N\,a_k(u) \psi_k(\eta)
\end{equation}
where $N$ is the truncation order, $a_k(u)$ are the unknown modes and 
$\psi_k(\eta)$ represent the basis functions. The field equations (\ref{eq:beta}), (\ref{eq:V}) and (\ref{eq:SWE}) are reduced to a set of ordinary differential equations for the modes $a_k(u)$. 

\subsection{Scripting}
In order to make a large amount of numerical experiments, minimize error of handling, processing the collected data, and for exploration near the critical point of bifurcation, we search for the critical point. Using  scripting with Python to ensure that the runs offered a mostly uniform coverage of the $\ln{(\lambda-\lambda^*)}$ space.

\section{Numerical experiments}
We explore numerically the power spectrum when a black hole forms
(super critical case).
Are there critical phenomena?
In the threshold of the black hole formation, is the spacetime discretely self-similar? 

Data A and C were used in Ref. \cite{bglw96}; they are not the best initial setting to explore in practice critical behavior. These data evolve too slow to reach critical values with the highest resolution, but they lead us to data C and D. However we extract an expected feature from datum A, that is, the mass gap can be reduced to zero. Figure 2 displays the mass gap as a function of $\lambda$ for different values of $\alpha$. 
Figure 3 illustrates the mass gap for the initial datum D, which is representative of any studied initial condition, except datum B, as explained in the caption of Figure 1. Figure 4 shows the mass spectrum in the super critical case for data C and D. For this particular graph, we have considered the Galerkin-Collocation method to evolve datum C. 
Figure 5 displays the natural logarithm of the Bondi mass as a function of $\tau$ defined as
\begin{equation}
\tau=-\ln[(t^*-t)/t^*],
\end{equation}
where $t^*$ is the accumulated Bondi time which corresponds to the accumulated proper time referred to $r=R$. 
  
From numerical experimentation we infer the following mass scaling power law
\begin{equation}
\delta M=K\delta\lambda^{2\gamma} + f [K\delta\lambda^{2\gamma}],
\end{equation}
where $\delta M=M_{B}-M^*$, $\delta\lambda=\lambda-\lambda^*$, being $f$ a non trivial function of its argument. $M^*$ is the super critical mass limit which corresponds to the critical amplitude $\lambda^*$. 

In general our results show agreement between the Finite differences and the Galerkin collocation method.

To obtain one point for the black hole spectrum $(M_{BH},\lambda)$ it takes 65 minutes for a grid size of $N_x=15,000$ using a N1-standard-1 virtual machine on Google Compute Engine.

\section{Discussion} 
When the spacetime contains a black hole and the critical behavior is studied, no mass gap is apparent \cite{mc}, \cite{cd15}. Simply the black hole increases the mass by the accreting massless scalar field.  As it was pointed in Ref. \cite{bglw96}, when the scalar field undergoes gravitational collapse to form a horizon, some of the scalar energy is radiated to infinity and the remainder crosses the horizon and contributes to the final black hole mass. The mirror itself must fall into the horizon for otherwise it would continue to reflect the scalar field until all scalar energy were radiated to infinity. Near the critical strength, the sensitivity of the final mass is somewhat analogous to the critical behavior studied by Choptuik \cite{c93} except there is now a mass gap because the final black hole must have a mass larger than $R/2$ in order to contain the mirror. Figures 2 and 3 show the evidence of the expected mass gap as a rule, and no mass gap as an exception. {The black hole mass in the critical behavior of type II can be finite.}

{The critical behavior as observed in this work is not exactly as reported up to now.} The kink setting may contain more complexity, {and} further study is planned.
The spectrum of the variation of the Bondi mass $\delta M$ as a function of the natural logarithm of the variation of the parameter, as displayed in figure 4, shows a curious behavior related with the standard critical phenomena. In fact, we observe the presence of $\Delta$ and $\gamma$ {as originally reported by Choptuik}, but in the present case the mass scaling power law is $2\gamma$. {We do not know yet why is roughly twice the critical exponent of Choptuik. It is intriguing indeed for us, and we are reporting it as it was found.} Even more, very close to the critical point appear an unstable main mode, followed by an unstable second mode. We do not know up to now if with a better resolution  will appear other subsidiaries minor modes or if the second mode is the dissipation of the main mode because of numerical error. We have confidence that {these peculiar results} are not a numerical artifact, instead they are a consequence of nonlinearity near the critical values and the boundary conditions. {Clearly, the mismatch in the critical exponent is due to the presence of the kink, constructed by means of a perfect reflecting barrier, and this makes the system different from the one originally studied by Choptuik.} 
{Other authors also have found different critical exponents when boundary conditions for the same system changed \cite{cd15} or other sources and  dimensions are considered \cite{gm}, \cite{jgt15}. However, the system as we studied basically behaves as Choptuik discovered: the spacetime is discretely self-similar and has a power law mass scaling. We emphasize in this paper the expected mass gap conjectured in Ref. \cite{bglw96}, where the system was studied with other motivations departing from that of Choptuik. Thus, the mass gap was previously conjectured but not found until now, and finding it was the main driver for this paper. The “universal" critical exponent is respect to the initial data, as has been clarified in many works after the pioneering work of Choptuik \cite{gm}.} Finally the Bondi mass can be singled out as in figure 5 in a such way that the periodic decaying in the subcritical case is $\Delta/2$, as expected for a dependence of the Bondi mass as $\lambda^2$, {in agreement with Ref. \cite{psa05}. }

The kink setting may look artificial or idealized. It is a perfect reflector barrier and just to the interior the spacetime is Minkowskian, but it can be filled with a fluid and the radius of the mirror incorporated to the dynamics. A number of other settings can be less unrealistic and the kink will behave basically the same way. 
The next problem to study is the kink under the axial and reflection symmetry, particularly motivated by the present results. Although the subcritical zone in the present setting deserves a more accurate resolution. Work in this direction is in progress.
  
\acknowledgments{W.B. thanks to FAPERJ for the financial support, and hospitality at Departamento de Física Teórica, UERJ; also to Luis Rosales, Luis Herrera and Jeff Winicour for reading and comments.}

\end{document}